\documentclass[twocolumn]{aastex63}

\received{Auggust 1, 2019}
\revised{May 14, 2020}
\accepted{June 29, 2020}

\submitjournal{ApJ}

\shorttitle{IGR J17591-2342}

\begin{document}

\title{Updated Spin and Orbital Parameters and Energy Dependent Pulse Behaviors of the Accreting Millisecond X-ray Pulsar IGR J17591-2342}

\correspondingauthor{Yi Chou}
\email{yichou@astro.ncu.edu.tw}

\author[0000-0002-7604-1517]{Kaho Tse}
\affiliation{Graduate Institute of Astronomy, National Central University, Jhongli 32001, Taiwan}
\affiliation{School of Physics and Astronomy, Monash University, Victoria 3800, Australia}

\author[0000-0002-8584-2092]{Yi Chou}
\affiliation{Graduate Institute of Astronomy, National Central University, Jhongli 32001, Taiwan}

\author{Hung-En Hsieh}
\affiliation{Graduate Institute of Astronomy, National Central University, Jhongli 32001, Taiwan}

\begin{abstract}

  We present our works in updating the spin and orbital parameters for the newly discovered accreting millisecond X-ray pulsar (AMXP) IGR J17591-2342 through pulsar timing and analyzing its energy dependent pulse behaviors. The data being analyzed were collected by {\it Neutron Star Interior Composition ExploreR (NICER)} that observed this AMXP from August to October 2018. Using the pulse arrival time delay technique, more accurate spin and orbital parameters were evaluated. From the measured spin frequency derivative, it is estimated that the magnetic field of the neutron star in IGR J17591-2342 is approximately $4 \times 10^{8}$ G. Precise pulse profiles can be made using the updated spin and orbital parameters. The soft phase lag phenomenon that is usually seen in other AMXPs is also observed from $\sim$4 keV to 12 keV with a value of 0.06 cycles (1.14 $\mu$s). Additionally, the pulsed fractional amplitude increases from 1 to $\sim$5 keV and then decreases for higher energy bands. We found that these phenomena, as well as the energy spectrum, can be explained by the two-component model \citep{gie02,pou03} with a relatively strong blackbody component and an additional unpulsed disk blackbody component.

\end{abstract}


\section{Introduction} \label{sec:intro}

Soon after the discovery of millisecond pulsars from radio band in the early 80's, low mass X-ray binaries (LMXBs) have been considered as their progenitors because a sufficient amount of angular momentum could transfer through the accretion disk to spin-up the neutron stars and enable them to be millisecond pulsars during the evolution of LMXBs~\citep{alp82,rad82}. Measuring the coherent pulsation in an LMXB provides an evolutionary link between radio millisecond pulsars and LMXBs. However, no coherent millisecond pulsation was detected in LMXBs until the discovery of the first accreting millisecond X-ray pulsar (AMXP), SAX J1808.4-3658, with a spin period of 2.5 ms~\citep{wij98,cha98}. To date, more than 20 LMXBs that have been identified as AMXPs~\citep{cam18} with pulsation periods ranging from 1.67 ms~\citep[IGR J00291+5934,][]{gal05} to 9.51 ms~\citep[IGR J16597-3704,][]{san18a}.     

The transient X-ray source IGR J17591-2342 was discovered by the INTernational Gamma-Ray Astrophysics Laboratory ({\it INTEGRAL}) on 2018 August 10 when it was scanning the Galactic center \citep{duc18}. The follow-up observations were then made in X-ray \citep{boz18,fer18,now18}, optical \citep{rusa18}, near-infrared \citep{sha18} and radio bands \citep{rusb18}. The Australia Telescope Compact Array (ATCA) made the observations on 2018 August 14, 19, and 25. In addition to reporting the precise source location, \citet{rus18} found that IGR J17591-2342 is a radio loud system in comparison to other accreting neutron stars in their outburst states although no radio pulsation was detected. The Chandra made a 20 ks follow-up observation on 2018 August 23 \citep{now18}. The spectrum detected by the High Energy Transmission Gratings Spectrometer indicates that there is an outflow with a velocity of $\sim$2800 km s$^{-1}$ from the system and the discovery of the overabundance of calcium implies that IGR J17591-2342 could be formed from the collapse of a white dwarf system  in a calcium-rich Type Ib supernova \citep{now19}. \citet{kri18} pointed out that the source had been detected by Swift/BAT on 2018 July 22, 20 days before its discovery by {\it INTEGRAL} and the flux reached to its peak on July 25. \citet{sanc18} and \citet{ray18} reported that the source experienced two rebrightenings starting on August 18 (MJD 58348) and September 6 (MJD 58366), respectively. \citet{fer18} identified that IGR J17591-2342 is an AMXP with a pulsation frequency of 527 Hz ($P_s \simeq 1.9$ ms), orbital period of 0.37 d ($\sim$32000 s), and a projected semi-major axis of 1.23 lt-s. More extensive studies was conducted by \citet{san18b} using the data collected by {\it Nuclear Spectroscopic Telescope Array (NuSTAR)}  on 2018 August 12 and {\it Neutron Star Interior Composition ExploreR (NICER)} from 2018 August 15 to 24. In addition to deriving more precise spin and orbital parameters of the system, the broadband X-ray spectral analysis  combining  {\it SWIFT} and {\it INTEGRAL} data was also made.

The primary scientific objective of {\it NICER} is to investigate the interior composition of neutron stars by constraining their masses and radii using the precise measurement of the pulse profiles but AMXPs are not suitable for this goal due to their complex accretion flows \citep{gen12}. Nevertheless, the pulse profiles of the AMXPs can give us clues to understand the natures of the systems, such as the emission mechanism of the pulsations. To obtain the precise pulse profile depends on accurate spin and orbital parameters for an AMXP. In this paper, we present our analysis results of timing and pulse profile properties of IGR J17591-2342 using the data collected by {\it NICER} of its 2018 outburst. The {\it NICER} observations and data reductions are briefly introduced in section~\ref{obs}. In section~\ref{dar}, we demonstrate the data analyses and results, including updating the spin and orbital parameters (section~\ref{ta}), the energy dependent pulse behaviors (section~\ref{pp}) and the spectral analysis (section~\ref{prs}). Finally, the analysis results, including the implications of the neutron star magnetic field from the measured spin frequency derivative and the energy dependent pulse behaviors, are discussed in section~\ref{dis}.

\section{Observations} \label{obs}

\begin{figure}
\plotone{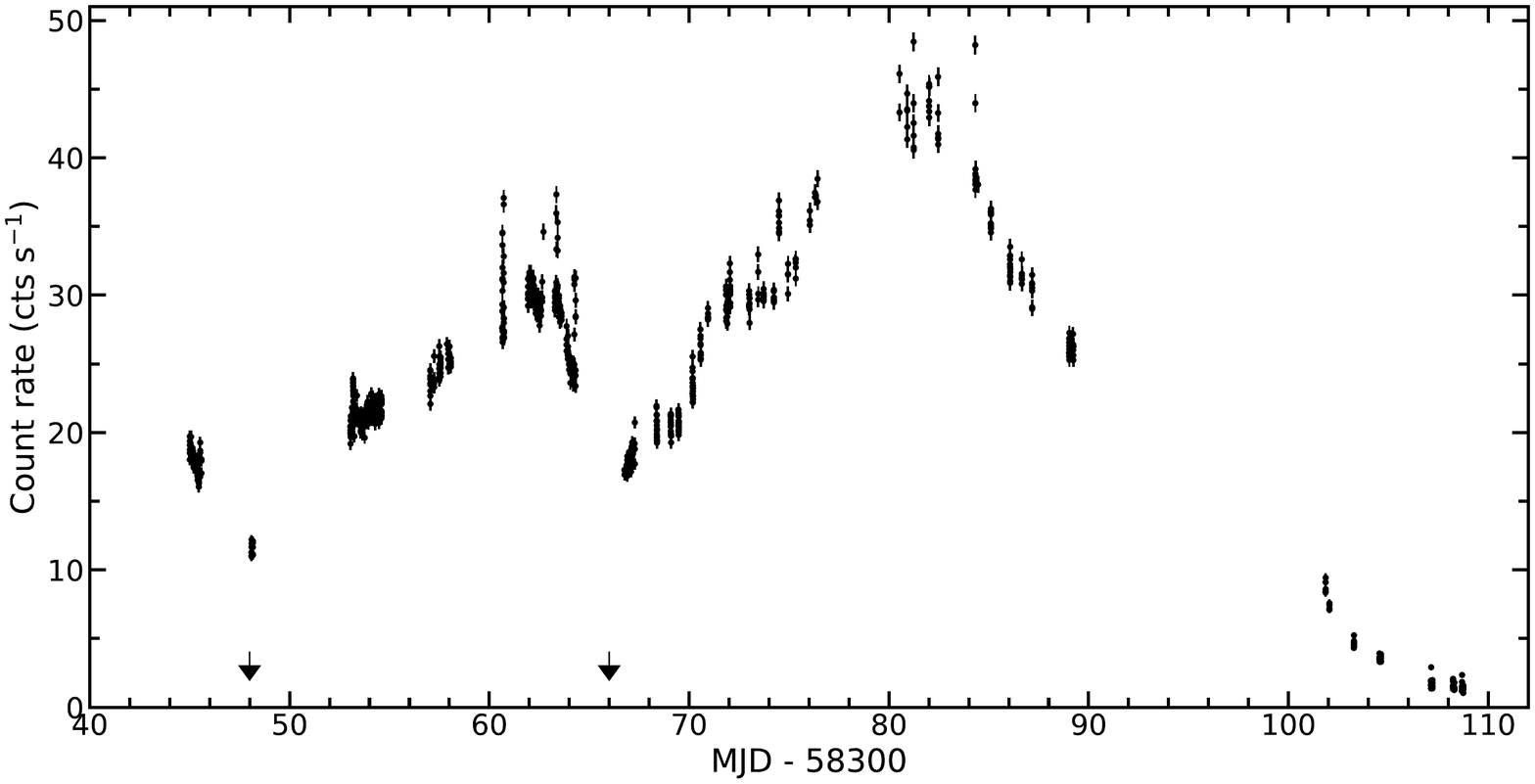}
\caption{The 1-12 keV X-ray light curve detected by {\it NICER} with a bin size of 100 s. The arrows indicate the start times of two re-brightenings on MJD 58348 and MJD58366. No pulsation can be detected from the data collected after MJD58400 (i.e. 2018 October observations) so these data were excluded from the analysis of this work.\label{lc}}
\end{figure}

{\it NICER} started observing the IGR J17591-2342 on 2018 August 14 (ObsID:1200310101), soon after an outburst alert of the source was issued \citep{duc18}, until near the end of the outburst on 2018 October 17 (ObsID:1200310139). The data files being analyzed in this work were first processed by the standard data filtering criteria described in the NICER Mission Guide\footnote{\url{https://heasarc.gsfc.nasa.gov/docs/nicer/mission_guide/}} with {\it NICERDAS} version 4.0 and the intervals with unexpected background flares were also excluded. Using the tool {\it barycorr} version 2.1 and by applying the JPL solar planetary ephemeris DE430, we corrected all the photon arrival times of events to the barycenter of the solar system according to the source position of the radio counterpart obtained by the ATCA observations \citep{rusb18}. To reduce possible noise and background, only the events between energy range 1 to 12 keV (PI values between 100 and 1200) with both slow and fast chains triggered were selected for {\bf timing} analysis. Figure~\ref{lc} shows that the  light curve of IGR J17591-2342 detected by {\it NICER} 1-12 keV band with the total exposure time of $\sim$91.5 ks. Two rebrightenings can be clearly seen in the light curve. More than the light curve shown in \citet{ray18}, X-ray flux reaches to its peak again  around 2018 September 19 to 23 (MJD 58380 to 58384) in the second rebrightening and then decays close to zero in the following $\sim$20 d. No type-I burst was detected for the whole {\it NICER} observations.

\section{Data Analysis and Results} \label{dar}
\subsection{Timing Analysis}\label{ta}

To update the spin and orbital parameters, all the selected events as described in section~\ref{obs} were divided into $\sim$300-sec data segments for subsequent analysis. To choose the data segments with significant pulsation detection, we made the power spectra for all data segments using $Z^2_1$ test \citep{buc83} based on the circular orbital parameters proposed by \citet{san18b} derived from the {\it NICER} observations made between 2018 August 15 and 24. Only the data segments with more than 99.9877\% confidence level {\bf ($Z^2_1 > 18$)} in pulsation detection were selected for further analysis. We found that all the data segments observed in August and September have significant pulsation detections but no pulsation can be found in the October observations, probably due to the very low count rates. Totally 264 data segments were selected for the following analysis.

To refine the spin and orbital parameters, a method similar to the one proposed by \citet{san16} was adopted. Suppose the orbital is circular, the relation between photon emission and arrival times, using the guess orbital parameters, can be written as 

\begin{equation}\label{e_corr}
  t^{\prime}=t-A^{(c)} \sin \left[2 \pi f_{orb}^{(c)} (t^{\prime}-T_{nod}^{(c)})\right]
\end{equation}
\noindent where $t^{\prime}$ is the photon emission time, $t$ is the photon arrival time, $A \equiv a_x\sin i/c$ where $a_x\sin i$ is the projected orbital radius, $f_{orb}$ is orbital frequency, $T_{nod}$ is the time of passage of ascending node, and the parameters with superscript (c) represent the guess (calculated) parameters. The constant time delay between the barycenters of binary and solar system, $D/c$, is ignored in Eq~\ref{e_corr}. The photon emission time $t^{\prime}$ can be solved numerically using the iteration method described in \citet{san16}.

For a constant spin frequency pulsar, the cycle count of the pulsar can be evaluated as
\begin{equation}\label{e_cycc}
N_c=\nu_0^{(c)}(t^{\prime}-T_0)
\end{equation}  
\noindent where $\nu_0^{(c)}$ is the guess spin frequency and $T_0$ is an arbitrary constant reference epoch. If there are small deviations between true and guess parameters, taking the first order approximation, the pulse phase drift can be described as

\begin{eqnarray}\label{e_phd}
 \delta \phi (t^{\prime}) & \approx & -\left({\partial N_c} \over {\partial \nu_0}\right )\Bigg |_{(c)} \delta \nu_0 -\left({\partial N_c} \over {\partial A}\right )\Bigg |_{(c)} \delta A \\
\nonumber & & -\left({\partial N_c} \over {\partial f_{orb}}\right )\Bigg |_{(c)} \delta f_{orb} -\left({\partial N_c} \over {\partial T_{nod}}\right )\Bigg |_{(c)} \delta T_{nod} 
\end{eqnarray} 
\noindent where $\delta$ are the differences between true and guess parameters, for example, $\delta \nu_0=\nu_0-\nu_0^{(c)}$. For the orbital parameter parts in Eq~\ref{e_phd}, the evaluated photon emission time $t^{\prime}$ from Eq~\ref{e_corr} is a function of guess orbital parameters, that is, $t^{\prime}=t^{\prime} (A^{(c)},f_{orb}^{(c)},T_{nod}^{(c)})$. Using Eq~\ref{e_cycc}, the partial differential on the right hand side of Eq~\ref{e_phd}, taking the parameter $A$ as an example, should be 

\begin{equation}\label{e_para}
\left({\partial N_c} \over {\partial A}\right )\Bigg |_{(c)}=\nu_0^{(c)}\left({{\partial t^{\prime}} \over {\partial A}}\right )\Bigg |_{(c)} 
\end{equation} 

\noindent and from Eq~\ref{e_corr}

\begin{eqnarray}\label{e_tpdif1}
  \left({{\partial t^{\prime}} \over {\partial A}}\right )\Bigg |_{(c)} &=&-\sin \left[2 \pi f_{orb}^{(c)} (t^{\prime}-T_{nod}^{(c)})\right]\\
\nonumber  & &-2\pi f_{orb}^{(c)}A^{(c)}\cos \left[2 \pi f_{orb}^{(c)} (t^{\prime}-T_{nod}^{(c)})\right]\left({{\partial t^{\prime}} \over {\partial A}}\right )\Bigg |_{(c)}  
\end{eqnarray} 

\noindent Therefore, solving (${\partial t^{\prime}}/ {\partial A}$) in Eq~\ref{e_tpdif1} and substituting it to Eq~\ref{e_para}, we obtain

\begin{equation}\label{e_tpdif2}
  \left({{\partial N_c} \over {\partial A}}\right )\Bigg |_{(c)} =-{{\nu_0^{(c)} \sin \left[2 \pi f_{orb}^{(c)} (t^{\prime}-T_{nod}^{(c)})\right]}\over {1+2\pi f_{orb}^{(c)}A^{(c)}\cos \left[2 \pi f_{orb}^{(c)} (t^{\prime}-T_{nod}^{(c)})\right]} }
\end{equation}

Similar derivation can be applied to the other orbital parameters, $f_{orb}$ and $T_{nod}$, and the pulse phase evolution can be expressed as

\begin{eqnarray}\label{e_tpdif3}
  \nonumber \phi(t_i) &=& \phi_0 -(t_i-T_0) \delta \nu \\
  \nonumber  &+& {{\nu_0^{(c)} \sin \left[2 \pi f_{orb}^{(c)} (t_i-T_{nod}^{(c)})\right]}\over {1+2\pi f_{orb}^{(c)}A^{(c)}\cos \left[2 \pi f_{orb}^{(c)} (t_i-T_{nod}^{(c)})\right]} } \delta A \\
  \nonumber  &+& {{2 \pi A^{(c)}\nu_0^{(c)} (t_i-T_{nod}^{(c)}) \cos \left[2 \pi f_{orb}^{(c)} (t_i-T_{nod}^{(c)})\right]}\over {1+2\pi f_{orb}^{(c)}A^{(c)}\cos \left[2 \pi f_{orb}^{(c)} (t_i-T_{nod}^{(c)})\right]} } \delta f_{orb} \\
  \nonumber  &-& {{2 \pi A^{(c)} f_{orb}^{(c)} \nu_0^{(c)} \cos \left[2 \pi f_{orb}^{(c)} (t_i-T_{nod}^{(c)})\right]}\over {1+2\pi f_{orb}^{(c)}A^{(c)}\cos \left[2 \pi f_{orb}^{(c)} (t_i-T_{nod}^{(c)})\right]} } \delta T_{nod} \\
\end{eqnarray} 

\noindent where $t_i$ is the time of ith data segment and $\phi_0$ is a constant phase that depends on the reference epoch $T_0$. By fitting the evolution of pulse phases with Eq~\ref{e_tpdif3}, the refined parameters can be derived by the corrections of the parameters. Taking the refined parameters as guess parameters, this correcting process can be iterated until the corrections of parameters become insignificant (e.g. smaller than the corresponding uncertainties of parameters).

Considering the spin frequency derivative, $\dot \nu$, Eq~\ref{e_cycc} should be rewritten as $N_c=\nu_0^{(c)}(t^{\prime}-T_0) + 1/2 \dot \nu^{(c)} (t^{\prime}-T_0)^2$ and the first order approximation of the pulse phase drift should be written as

\footnotesize
\begin{eqnarray}\label{e_tpdif4}
\nonumber &\phi(t_i)& = \phi_0 -(t_i-T_0) \delta \nu -{1 \over 2}  (t_i-T_0)^2 \delta \dot \nu\\
  \nonumber  &+& {{\left[\nu_0^{(c)}+\dot \nu^{(c)}(t_i-T_0)\right ] \sin \left[2 \pi f_{orb}^{(c)} (t_i-T_{nod}^{(c)})\right]}\over {1+2\pi f_{orb}^{(c)}A^{(c)}\cos \left[2 \pi f_{orb}^{(c)} (t_i-T_{nod}^{(c)})\right]} } \delta A \\
  \nonumber  &+& {{2 \pi A^{(c)} \left[\nu_0^{(c)}+\dot \nu^{(c)}(t_i-T_0)\right](t^{\prime}-T_{nod}^{(c)}) \cos \left[2 \pi f_{orb}^{(c)} (t_i-T_{nod}^{(c)})\right]}\over {1+2\pi f_{orb}^{(c)}A^{(c)}\cos \left[2 \pi f_{orb}^{(c)} (t_i-T_{nod}^{(c)})\right]} } \delta f_{orb} \\
  \nonumber  &-& {{2 \pi A^{(c)} f_{orb}^{(c)}  \left[\nu_0^{(c)}+\dot \nu^{(c)}(t_i-T_0)\right]\cos \left[2 \pi f_{orb}^{(c)} (t_i-T_{nod}^{(c)})\right]}\over {1+2\pi f_{orb}^{(c)}A^{(c)}\cos \left[2 \pi f_{orb}^{(c)} (t_i-T_{nod}^{(c)})\right]} } \delta T_{nod} \\
\end{eqnarray}
\normalsize

\noindent To allow small eccentricity in the orbital model, two additional terms, $({e\cos\omega}/2) A^{(c)} [\nu_0^{(c)}+\dot \nu^{(c)}(t_i-T_0)]\sin [4 \pi f_{orb}^{(c)} (t_i-T_{nod}^{(c)})]$ and $-({e\sin \omega}/2) A^{(c)} [\nu_0^{(c)}+\dot \nu^{(c)}(t_i-T_0)]\cos [4 \pi f_{orb}^{(c)} (t_i-T_{nod}^{(c)})]$, are added into Eq~\ref{e_tpdif4} for the fitting, where $e$ is eccentricity and $\omega$ is the longitude of the periastron, .

\begin{figure}
\plotone{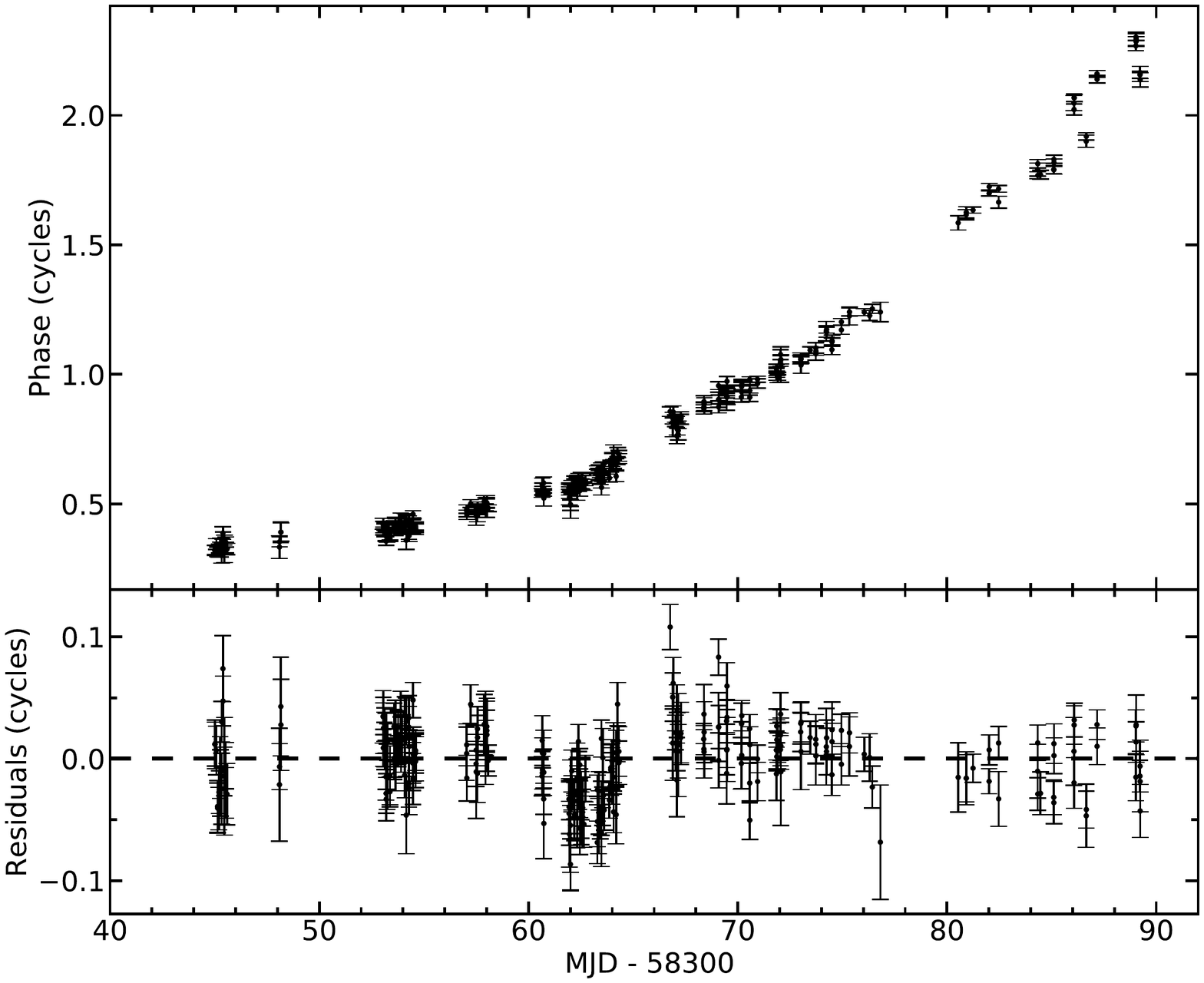}
\caption{The pulse phase residuals by applying the spin and orbital parameters proposed by \citet{san18b}, including $\dot \nu$ (=$2.0 \times 10^{-13}$ Hz s$^{-1}$) (top), and the updated parameters of Model 1 in Table~\ref{para} (bottom).\label{phase}}
\end{figure}

The event times of each selected data segment were first folded with the initial spin and circular orbital parameters proposed by \citet{san18b}, evaluated from 2018 August 15-24 {\it NICER} observations, and the phases were subsequently binned into 32 bins for a cycle to obtain the pulse profile. We found that all the pulse profiles can be well-fitted with a two-component sinusoidal function, that is $r(\phi)=a_0+ \sum\limits_{k=1}^2[a_k\cos(2 \pi k \phi)+b_k\sin(2 \pi k \phi)]$. The peak of the best fitted profile was selected as the fiducial point of the pulse phase for each data segment. The error of the pulse phase was evaluated using $10^3$ runs of Monte Carlo simulation. If the initial parameters were correct, the pulse phase should have been consistent with being a constant within the measurement errors. The pulse phase evolution folded by the initial guess parameters is shown as Figure~\ref{phase}. A long-term trend of phase evolution can be clearly seen, that indicates that the spin and orbital parameters have to be further refined.

The refining procedure as described above was adopted using the model containing spin ($\nu$ and $\dot \nu$) and orbital parameters (hereafter Model 1) to update the parameters by selecting $T_0$ on MJD 58338 that is close to the  arrival time of the first event of {\it NICER} observations. Figure~\ref{phase} shows the pulse phase residuals before and after completion of the refining process and the updated spin and orbital parameters obtained by {\it NICER} observations are listed in Table~\ref{para}. The errors of spin parameters were quadratically added with the errors due to uncertainties of the source position proposed by \citet{rusb18}. A clear spin frequency derivative with a spin-down rate of $(-7 \pm 1) \times 10^{-14}$ Hz s$^{-1}$ was detected from {\it NICER} 2018 August to September observations. The eccentricity parts were further added into the model and found no significant eccentricity can be detected with 2$\sigma$ upper limit of $1 \times 10^{-5}$. 

\begin{figure}
\plotone{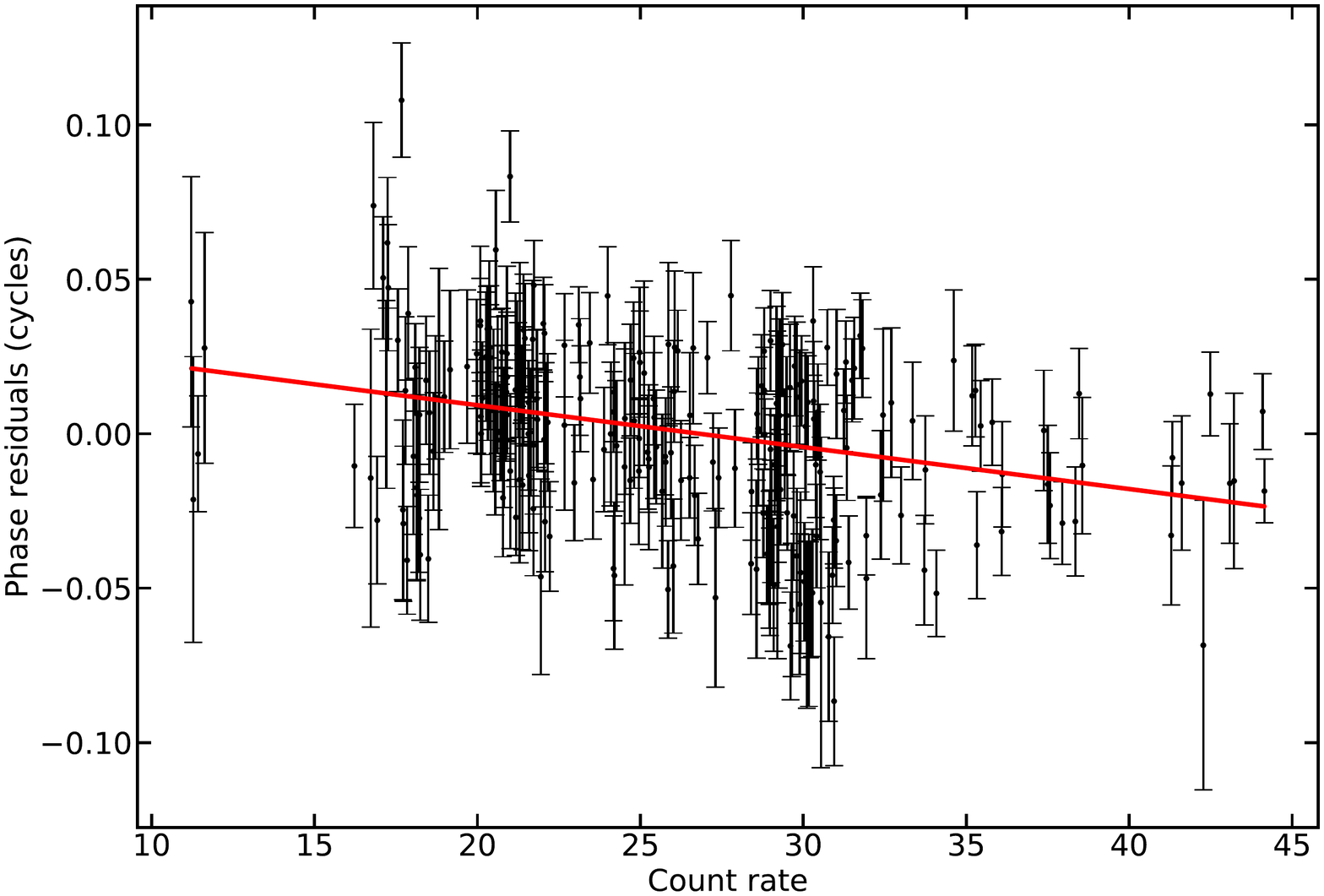}
\caption{The correlation between pulse phase residuals folded by the Model 1 in Table~\ref{para} and the corresponding 1-12 keV count rates detected by {\it NICER}. The red line is the best fit of a linear function\label{phcts}}
\end{figure}

\begin{figure}
\plotone{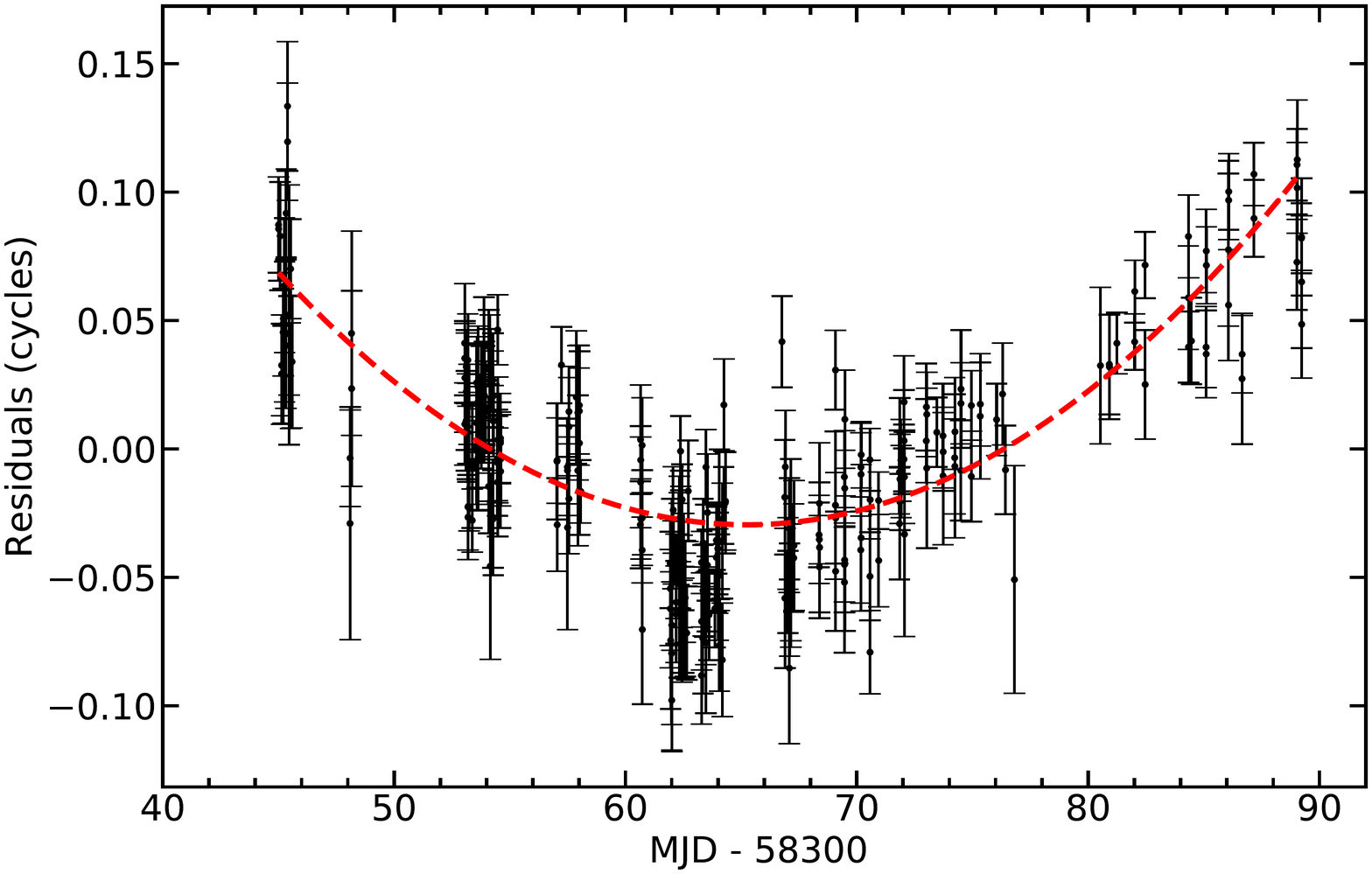}
\caption{The quadratic trend of pulse phase residuals folded by a constant $\nu$ plus a Keplerian orbit model with the orbital parameters and $\alpha$ fixed at the values list in Table~\ref{para} {\bf (Model 2)}. The red dashed line is the best-fit quadratic function for the pulse phase evolution with $\dot \nu= (-6 \pm 1) \times 10^{-14}$ Hz s$^{-1}$. \label{nudot}}
\end{figure}

However, the fitting from the Model 1 gives a large reduced $\chi^2$ value of 2.43 for 258 degree of freedom (d.o.f.) that is not statistically acceptable. Such large deviation may be caused by the timing noise due to the movement of the hot spot on the neutron star surface that is highly correlated with X-ray flux observed in many of AMXPs~\citep{pat09a}. Furthermore, this phase drift may also induces spurious spin frequency derivative in detection~\citep{pat09a}. To verify if this flux-dependent timing noise makes the large deviations, we first evaluated the linear Pearson correlation coefficient between the pulse phase residuals and the corresponding X-ray count rates. A correlation coefficient of -0.34 for 264 data points with a p-value of $1.3 \times 10^{-8}$ was obtained that implies the high correlation between the pulse phase residuals and the X-ray count rates (Figure~\ref{phcts}). To minimize the effect from the flux-dependent timing noise, we adopted the numerical simulation results from~\citet{kul13} that the azimuthal location of the hot spot is proportional to ${\dot M}^{-1/5}$ where $\dot M$ is the accretion rate~\citep[also see][]{bul19a}. Assuming that the 1-12 keV count rate is about proportional to the accretion rate, the phase drift due to the hot spot movement can be written as $\Delta \phi_n=\alpha{r}^{-1/5}$ where $r$ is count rate and $\alpha$ is a constant. We therefore added an additional term, $-\alpha{r_i}^{-1/5}$ where $r_i$ is the count rate of ith data segment, to the Eq~\ref{e_tpdif4}. The optimal spin and orbital parameters, as well as the proportional constant $\alpha$, for this revised model (hereafter Model 2) are listed in Table~\ref{para}. The spin frequency derivative $\dot \nu$ changes from $(-7 \pm 1) \times 10^{-14}$ Hz s$^{-1}$ to $(-6 \pm 1) \times 10^{-14}$ Hz s$^{-1}$, that implies the $\dot \nu$ evaluated from the Model 1 is likely partly attributed to the flux-dependent timing noise. Considered together with this timing noise component, Figure~\ref{nudot} shows quadratic trend of the pulse phase residuals folded by the constant spin frequency model plus the Keplerian orbit with the orbital parameters and $\alpha$ fixed at their optimal values listed in Table~\ref{para} (Model 2). Although the reduced $\chi^2$ value decreases significantly from 2.43 to 1.85, that indicates the revised model is better than the previous one, the value is still too large to be statistically acceptable. However, there is no structure in the residuals. This poor fitting is likely due to the remaining unmodelled timing noise~\citep{pat12}.

 On the other hand, we cannot exclude the possibility that the quadratic trend in the pulse phase evolution is due to red noise. To verify this, a simulation of a first order autoregressive (AR(1)) process as $\phi_i=a_i \phi_{i-1}+ \varepsilon_i$ was made. First, assuming that the quadratic trend on Figure~\ref{nudot}  (with $\dot \nu= -6 \times 10^{-14}$ Hz s$^{-1}$) is purely due to the AR(1) process, we selected the equal-spacing points from the curve and evaluated $a_i=a=constant$ using the autocorrelation function. Further suppose that the process is damped random walk, $a=exp(-\Delta t/\tau)$ where $\Delta t$ is the time interval of the two consecutive equal-spacing points. The damping timescale $\tau$ can be obtained. Applying the $\tau$ to the unevenly-spacing data points of pulse phase evolution, we simulated an AR(1) process as $\phi_i=exp[-(t_i-t_{i-1})/\tau]\phi_{i-1}+\varepsilon_i$ where $t_i$ is the time of ith data segment and $\varepsilon_i$ is pure Gaussian white noise with mean zero and the variance that equals to the square of phase error of ith data segment, and then fitted a quadratic curve to the simulated data to find the probability that $|\dot \nu|> 6 \times 10^{-14}$ Hz s$^{-1}$ and $\chi^2_\nu<1.85$. After $10^6$ runs of the Monte Carlo simulation, none of them is satisfied the criteria above. We therefore conclude that the quadratic trend is unlikely caused by the red noise with a probability less than $10^{-6}$.

\subsection{Energy Dependent Pulse Behaviors}\label{pp}

\begin{figure}
\plotone{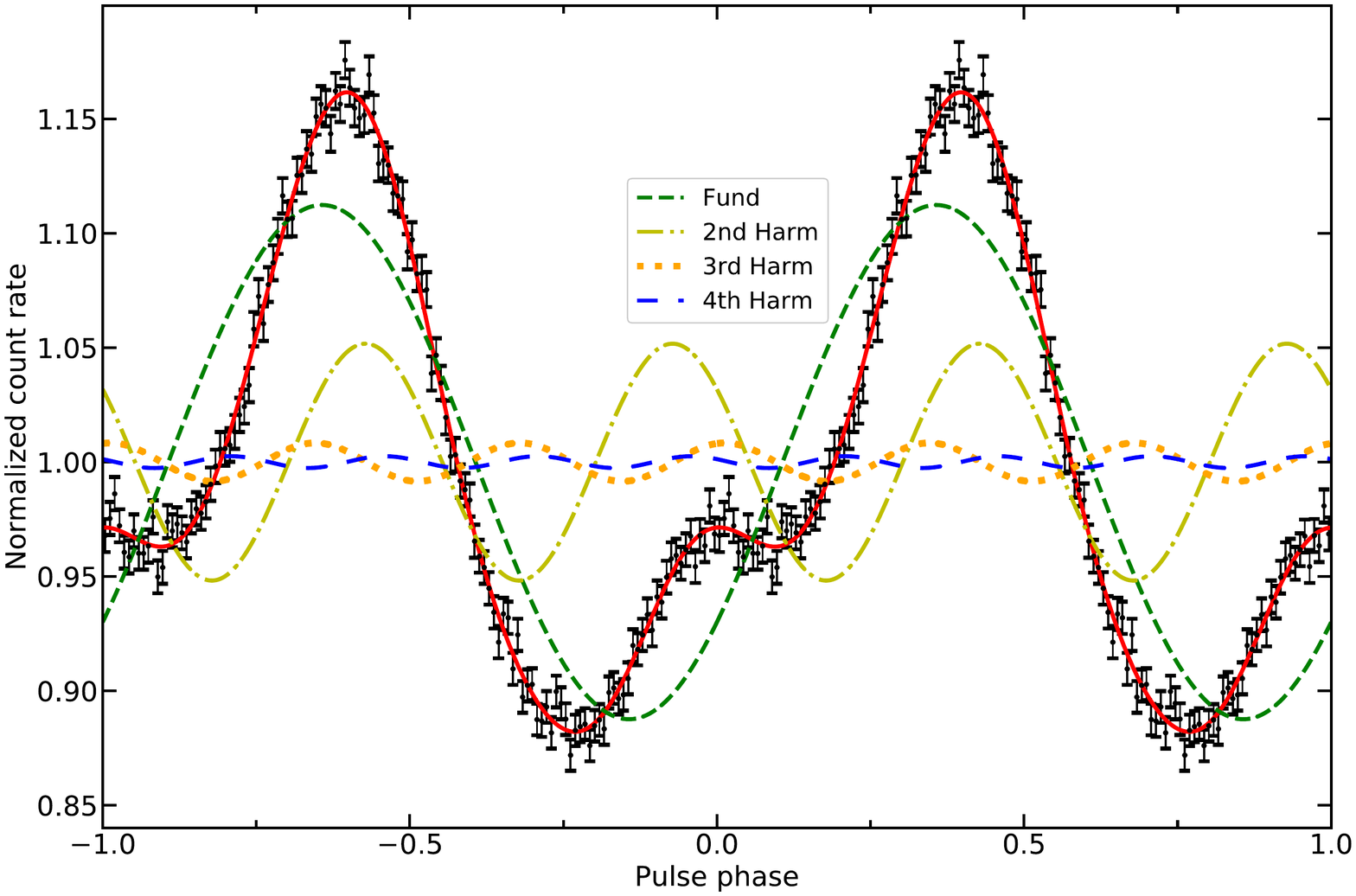}
\caption{The pulse profile made by folding the events of {\it NICER} 2018 August and September observations with the {\bf optimal} spin and orbital parameters listed in Table~\ref{para}. It needs a four-component sinusoidal function (fundamental+3 overturns, red solid line) to describe it. The fundamental, second, third and fourth components are represented in green dashed, yellow dash-dotted, orange dotted and blue long-dashed lines, respectively. \label{profileall}}. 
\end{figure}

The combined pulse profile of all the data segments with significant pulse detections folded with the optimal spin and orbit parameters listed in Table~\ref{para} is shown as Figure~\ref{profileall} with a root mean square (rms) pulsed fractional amplitude of 8.8\%. Fitting the multiple sinusoidal functions to the pulse profile shows that it needs a four-component sinusoidal function, that is, $r(\phi)=a_0+ \sum\limits_{k=1}^4[a_k\cos(2 \pi k \phi)+b_k\sin(2 \pi k \phi)]$, to obtained an acceptable fitting with $\chi^2_{\nu}=0.98$. The rms pulsed fractional amplitude of the four components are 7.95\%, 3.66\%, 0.60\% and 0.18\% for the fundamental, second, third and fourth harmonics, respectively.

\begin{figure}
\plotone{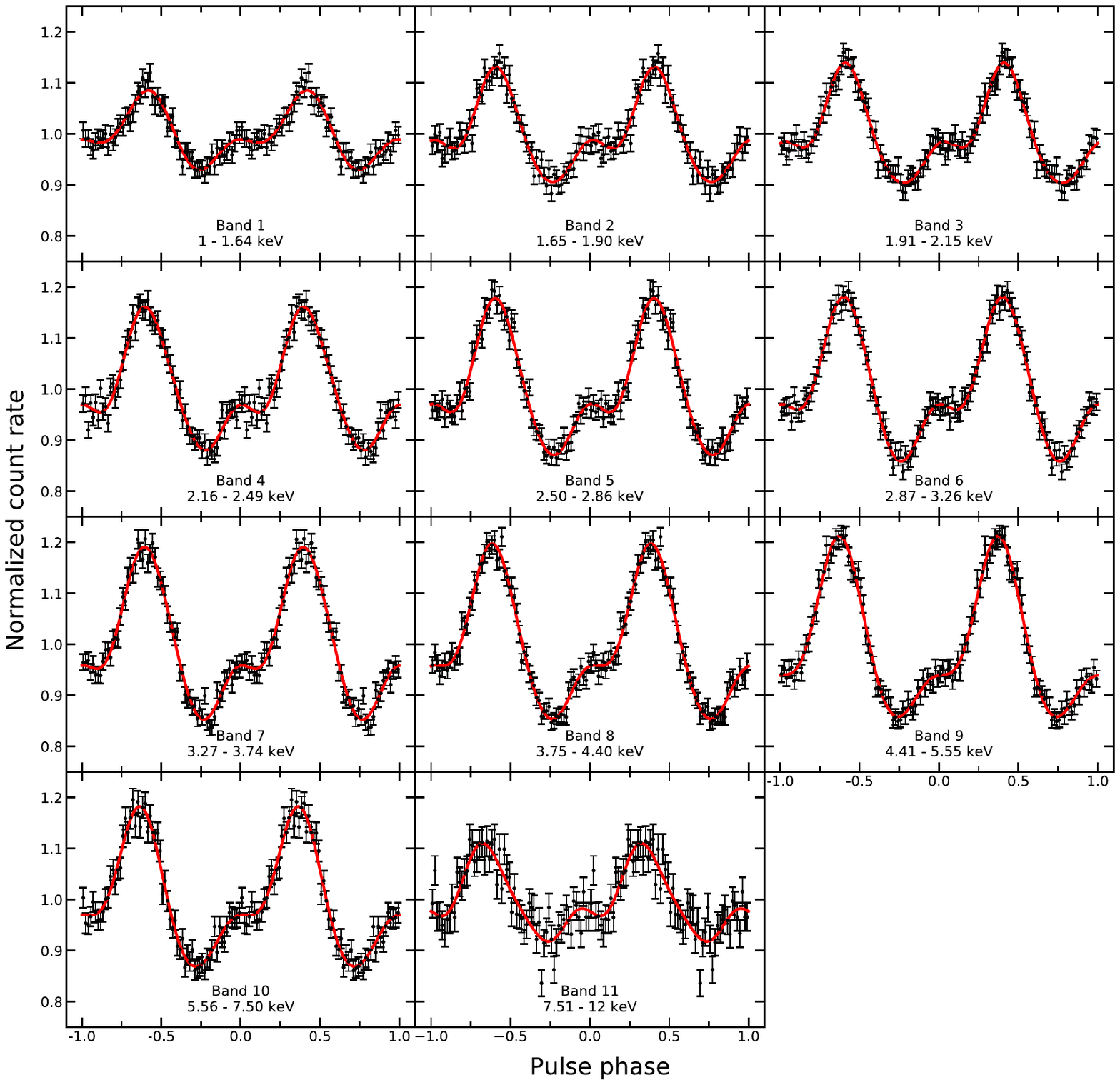}
\caption{The pulse profiles of 11 energy bands. All the profiles are well-fitted with three-component sinusoidal functions \label{profileeb}}
\end{figure}

To investigate the energy dependent pulse behaviors, the 1 to 12 keV events were divided into 11 energy bands. Each of them has same event numbers roughly, except for the band 10 and 11 that have only about half of the event numbers compared with the other bands for further investigation of the pulse behaviors of harder X-ray bands. The pulse profile of these 11 bands are shown as Figure~\ref{profileeb}. All pulse profiles are well-fitted with a three-component sinusoidal function. To study the energy dependent pulse arrival time delay, that have been seen in many other AMXPs~\citep{pat12}, we applied the cross-correlation between the best fitted pulse profiles and the one of the softest band (band 1, 1-1.64 keV). The uncertainties of the delays were evaluated by $10^3$ runs of Monte Carlo simulation, except for the band 1 that was directly used the uncertainty of its fiducial point. Figure~\ref{ade} shows the energy dependent pulse arrival delay relative to the band 1. There is no significant soft lag up to band 7 (3.74 keV) and pulsations of harder energy bands lead the band 1 up to  $\sim$0.06 cycles ($\sim$114 $\mu s$) at the hardest band (band 11, 7.51-12 keV). There is neither soft lag saturation \citep[e.g. HETE J1900.1-2455,][]{gal07} nor hard lags \citep[e.g. IGR J00291+5934,][]{fal05} can be seen in IGR J17591-2342 from the {\it NICER} observations. Figure~\ref{pe} shows the rms pulsed fractional amplitudes of different energy bands. They first monotonically increase from $\sim$4.7\% for band 1 up to $\sim$11.5\% for band 9 (4.41-5.55 keV) and then decrease to $\sim$5.9\% in the hardest energy band (band 11, 7.51-12 keV). It implies that there may be soft unpulsed emissions in the spectrum, that will be further discussed in Section~\ref{prs} and~\ref{dis}.

\subsection{Energy Spectrum}\label{prs}

The energy dependent pulse behaviors of an AMXP would be highly related to its energy spectrum (see Section~\ref{dis} for more detail discussions), so we performed spectral analysis of the data collected by {\it NICER}. The spectrum of energy range from 0.7 to 12 keV was extracted from the same dataset as the one used for timing analysis. The background was evaluated from {\it NICER} observations of Rossi X-ray Timing Explorer ({\it RXTE}) blank field region 5~\citep{jah06}, followed by the same filtering as we processed to the source data. To explain the energy dependent pulse arrival time delay detected in SAX J1808.4-3658, \citet{gie02} and \citet{pou03} proposed a two-component model composed of a blackbody emission from a hot spot and a Comptonized emission from a hot slab, to describe the pulsed emission from an AXMP. Using XSPEC version 12.10~\citep{arn96}, the spectrum was first modeled with an absorbed blackbody plus a Comptonized continuum as {\bf tbabs(bbodyrad+nthcomp)}. For the Comptonized component, we found it is hard to well constrain on the electron temperature that is likely due to the lower energy range (0.7-12 keV) for the observed spectrum. The electron temperature was therefore fixed at 22 keV, same as the one found in IGR J17591-2342 by~\citet{san18b} from the broad-band spectral fitting. However, a large reduced $\chi^2$ value of 3.13 (d.o.f.=1125) was yielded from the fitting. A majority deviation between the data and the model was found in the low energy range ($<$1.4 keV).

The energy dependent pulsed fractional amplitudes (Figure~\ref{pe}) imply that there could be soft, unpulsed component in the spectrum. Therefore, a multicolor disk blackbody component was added into the spectral model, that is, {\bf tbabs(diskbb+bbodyrad+nthcomp)}. Adding this additional component significantly improved the fitting with a reduced $\chi^2$ of 1.54 (d.o.f.=1123) and the spectral parameters are presented in Table~\ref{specpara}. Figure~\ref{spec} shows the unfolded spectrum of the {\it NICER} dataset we analyzed. Although this reduced $\chi^2$ is still statistically unacceptable, the large change of the $\chi^2$ ($\Delta \chi^2=1792$) implies that the unplused disk blackbody is an essential component for the spectral model. On the other hand, we noticed that the remaining residuals are mostly concentrated below $\sim$2.5 keV, that is likely caused by the instrumental residuals in the {\it NICER} response function. In addition to the disk blackbody component, it is evident that the emission from the blackbody component dominates the one from the Comptonized component for photon energies $\lesssim$ 3.2 keV. This characteristic is related to the energy dependent pulse arrival time that will be further discussed in Section~\ref{dis}.

\section{Discussion} \label{dis}

We have updated the spin and orbital parameters of AMXP IGR J17591-2342 from the data collected by {\it NICER} in 2018 August and September. The updated parameters are generally consistent with the those reported by~\citet{san18b} but with more precise values (see Table~\ref{para}). The only exception is the spin frequency derivative. We found that the neutron star in IGR J17591-2342 is, averagely speaking, spun-down with a rate of $-6 \times 10^{-14}$ Hz s$^{-1}$ during the entire duration of {\it NICER} observations, instead of being a constant spin frequency (with an insignificant spin-up rate of $(2.1 \pm 1.6) \times 10^{-13}$ Hz $s^{-1}$) as proposed by~\citet{san18b} from 2018 August 15 to 24 observations. This value essentially agrees with the theoretical prediction ($\sim 10^{-13}$ Hz s$^{-1}$). The spin frequency derivative allows us to roughly estimate the magnetic field on the surface of neutron star in this AMXP. The neutron star can be spun-up by accretion and be spun-down due to the magnetic drag in the accretion disk and the magnetic dipole radiation. Using Equation (23) in~\citet{pap04} for fast X-ray pulsars but with also considering the magnetic dipole radiation, the torque acting on the neutron star can be rewritten as 

\begin{equation}\label{torque}
\Gamma=2 \pi I \dot \nu \simeq \dot M \sqrt{GMr_c}-{\mu^2 \over 9r_c^3}-{{2} \over {3}} \mu^2 \sin^2 \alpha \Bigl({{2 \pi \nu} \over {c}}\Bigr)^3
\end{equation}   

\noindent where $I$ is the moment of inertia of neutron star, $\dot M$ is the mass accretion rate, $M$ is the neutron start mass, $\mu$ is the magnetic dipole moment, $r_c$ is the corotation radius that the radius with the Keplerian frequency is equal to the spin frequency, $\alpha$ is the angle between rotational axis and magnetic dipole, $G$ is the gravitational constant and $c$ is the speed of light. Using the numerical values of $\dot M \simeq 5.2 \times 10^{-10} M_{\sun}$ $yr^{-1}$ evaluated from the broad-band spectrum~\citep{san18b}, $M=1.4 M_{\sun}$, $r_c=(GM/4\pi^2 \nu^2)^{1/3}$=25.7 km, $I \simeq 10^{45}$ g cm$^2$, and $\nu$ and $\dot \nu$ from Table~\ref{para}, the magnetic dipole moment could be estimated as $\mu \simeq (3.8-4.1) \times 10^{26}$ G cm$^3$ for $\sin \alpha$=1 to 0, respectively, that implies that the magnetic field $B \simeq 4 \times 10^{8}$ G on the neutron star for a radius of 10 km, consistent with the magnetic field of $10^8-10^9$ G for millisecond pulsars. However, we note that the spin frequency derivative contributed by the magnetic dipole radiation can be as high as $-2.2 \times 10^{-14}$ Hz s$^{-1}$ (for $\sin\alpha=1$), $\sim 37\%$ of the total spin frequency derivative. If we assume that the magnetospheric radius $r_m \simeq r_c= \xi r_A$ where $r_A=(\mu^4/2GM\dot M^2)^{1/7}$ is the Alfv\'{e}n radius, form the numerical values above, $r_A \simeq$ 48.6 km was obtained and it indicates $\xi \simeq$0.53, close to the typical value for disk accretion ($\xi \sim 0.5$).

The energy dependent pulse arrival time can be usually observed in AMXPs. Soft lags are most often seen in AMXPs, where the pulse arrival times of the pulsations from softer energy bands lag compared to the ones from the harder energy bands. This phenomenon was first discovered by \citet{cui98} during the 1998 outburst of the SAX J1808.4-3658. The soft lag tendency can clearly be detected from 2 keV extended up to 10 keV (hereafter break point) for about 200 $\mu$s ($\sim$0.08 cycles) but saturated for harder X-ray bands. Similar energy dependent pulse arrival time behavior can be also found in XTE J1751-305 \citep{gie05}, XTE J1814-338 \citep{wat06}, HETE J1900.1-2455 \citep{gal07} and XTE J1807-294 \citep{chou08} with different break point energies. However, the energy dependent pulse arrival time acts differently in IGR J00291+5934. ~\citet{fal05} found that the soft lags can extends to the break point energy at $\sim$6 keV but the tendency reverses instead of saturation for the pulse of energy bands harder than the break point (hereafter hard lags) during its 2005 outburst. Similar behaviors are also observed during its 2015 outburst \citep{san17} with a different break point energy at $\sim$8 keV. Such hard lags are also marginally detected in IGR J17511-3057 \citep{fal11} and IGR J17498-2921 \citep{fal12}. On the other hand, the energy dependent pulse arrival times for two AMXPs behalf rather peculiarly. The pulse arrival times seem independent of energy bands below $\sim$17 keV for SAX J1748.9-2021~\citep{pat09b} and only hard lags are detected in IGR J18245-2452 \citep{def17}. The newly discovered AMXP IGR J17379-3747 also shows unusual energy dependent pulse arrival times. Regular soft lags can be observed between 0.5 and 6 keV with no break point being detected; however the phase difference between the softest and hardest bands can be as high as $\sim$0.3 cycles (or $\sim 110^\circ$) \citep{bul19b}.

A two-component model was proposed by \citet{gie02} and \citet{pou03} to explain the soft lags seen in SAX J1808.4-3658 during its 1998 outburst. There are two emission components, a soft blackbody component from a hot spot on the surface of neutron star around the accretion channel and a hard Compotonized component from up-scattering of the cooler seed photons by the hot electrons in a slab of accretion shock above the hot spot. Due to the scattering, the angular distribution of Comptonized photons is wider than that of the blackbody. Considered with Doppler boostering from the fast rotating AMXP, and the light bending in the Schwarzchild geometry, the pulsation from the Comptonized component leads to the pulsation from the blackbody component. For the softest energy band, the blackbody component dominates the emissions. As the energies of photons increased, the Comptonized emissions becomes stronger and the blackbody component becomes weaker. This causes the pulsation in harder energy bands to precede those in soft (i.e soft lags). Beyond the break point, the effect from the blackbody component can be completely neglected, leaving only the Comptonized component, so the energy dependent pulse arrival time reaches saturation. This two-component model can well explain the energy dependent pulse arrival times of SAX J1808.4-3658 \citep{cui98}, XTE J1751-305 \citep{gie05}, XTE J1814-338 \citep{wat06}, HETE J1900.1-2455 \citep{gal07} and XTE J1807-294 \citep{chou08}  but cannot be applied to hard lags seen in IGR J00291+5934~\citep{fal05,san17}.

\begin{figure}
\plotone{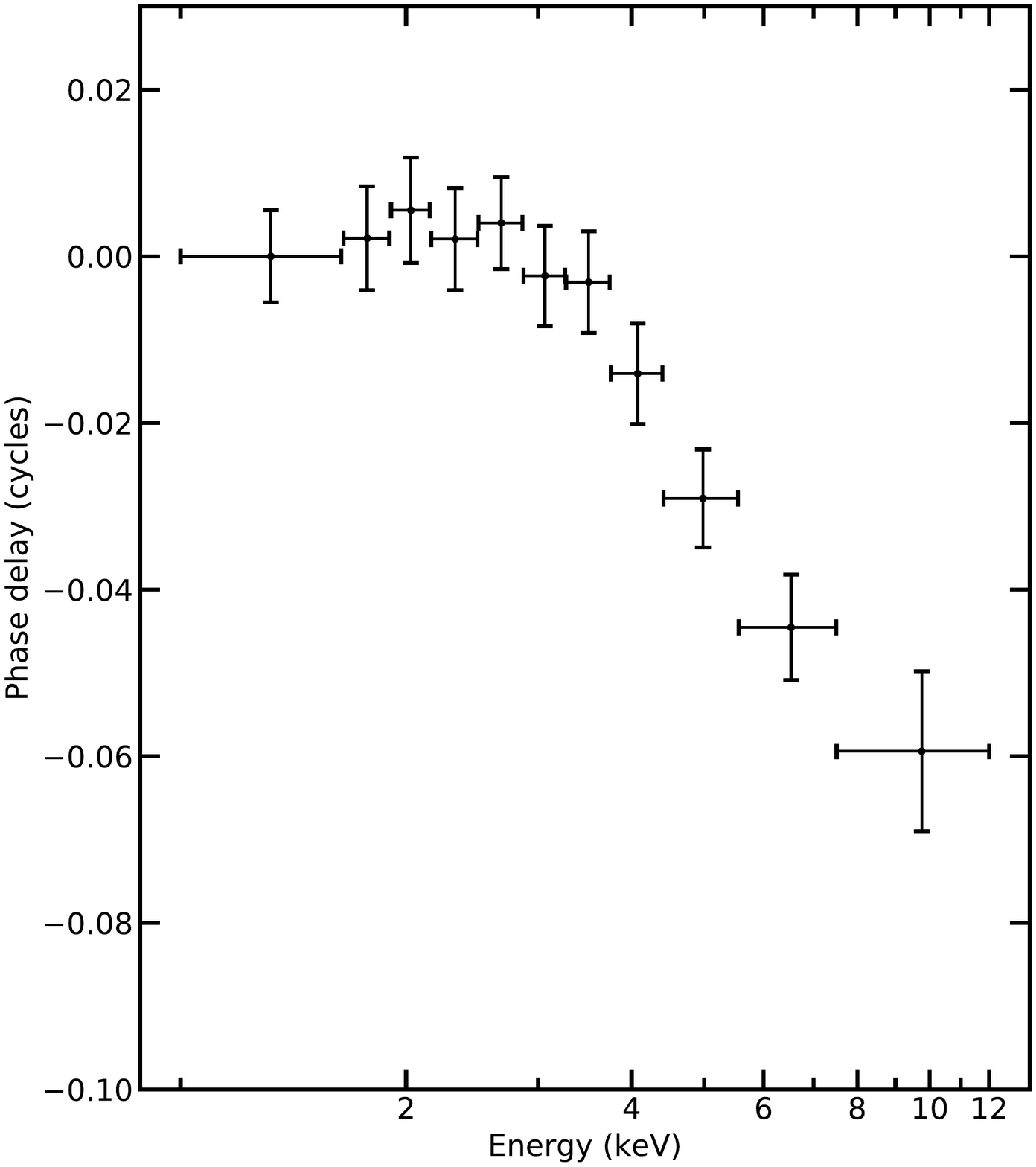}
\caption{Energy dependent pulse arrival time delay relative to the softest energy band (band 1, 1-1.64 keV). The negative values indicate that the pulse lead to the softest one.\label{ade}}
\end{figure}

We attempt to adopt the two-component model to explain the energy dependent behaviors of pulsations presented in section~\ref{dar}. Figure~\ref{ade} shows that the pulse phases are nearly constant between 1 to $\sim$4 keV and the soft lags appear for the harder energy bands. This indicates that the blackbody emissions from the hot spot dominates the Comptonized component's emission for the photon energy below $\sim$4 keV. The spectral analysis results also support this implication. The blackbody component is stronger than the Comptonized component for photon energy $\lesssim$4keV as shown in Figure~\ref{spec}. The blackbody flux is a factor of 1.6 (=8.14/5.13) larger than the Comptonized flux for 1-4 keV band even though the Comptonized flux is a factor of 2.3 (=21.5/9.42) larger than the blackbody flux for 1-12 keV band (see Table~\ref{specpara}). It indicates that the IGR J17591-2342 has a relatively strong blackbody emissions. Similar phenomenon can be observed in HETE J1900.1-2455 \citep[Figure 3]{gal07}, IGR J17511-3057 \citep[Figure 6,]{fal11} and IGR J17498-2921 \citep[Figure 5,]{fal12}. By contrast, some AMXPs show sharp soft lag tendency starting at $\sim$2 keV, such as SAX J1808.4-3658 in the 1998 outburst \citep[Figure 4]{cui98} and IGR J00291+5934 in 2005 outburst \citep[Figure 6]{fal05}, that implies that these sources may have relatively weaker blackbody emissions from the hot spot.

We did not detect the break point in IGR J17591-2342 from its 2018 outburst between 1 and 12 keV energy band. This implies that the break point is likely $\ga$10 keV. From the two component model, the break point energy indicates that the contribution from the blackbody component can be ignored beyond this energy. Some AMXPs have softer break point energies, such as XTE J1814-338~\citep[at $\sim$6 keV, see Figure 2 in][]{wat06}, HETE J1900.1-2455~\citep[at $\sim$6 keV, see Figure 3 in][]{gal07}, IGR J00291+5934 in the 2005 outburst~\citep[at $\sim$6 keV, see Figure 6 in][]{fal05} and 2015 outburst~\citep[at $\sim$8 keV, see Figure 4 in][]{san17} where as a few AMXPs have harder break point energies, such as XTE J1751-305~\citep[$>$10 keV, see Figure 8 in][]{gie05} and IGR J17511-3057~\citep[$>$10 keV, see Figure 6 in][]{fal11}. \citet{fal11} argued that the harder break point energy could be due to higher blackbody component temperature.  However, a similar effect could also be the result of a relatively strong blackbody emission that extends the contribution from the pulsation of the blackbody component to a harder energy band, as what we observed in IGR J17591-2342.

\begin{figure}
\plotone{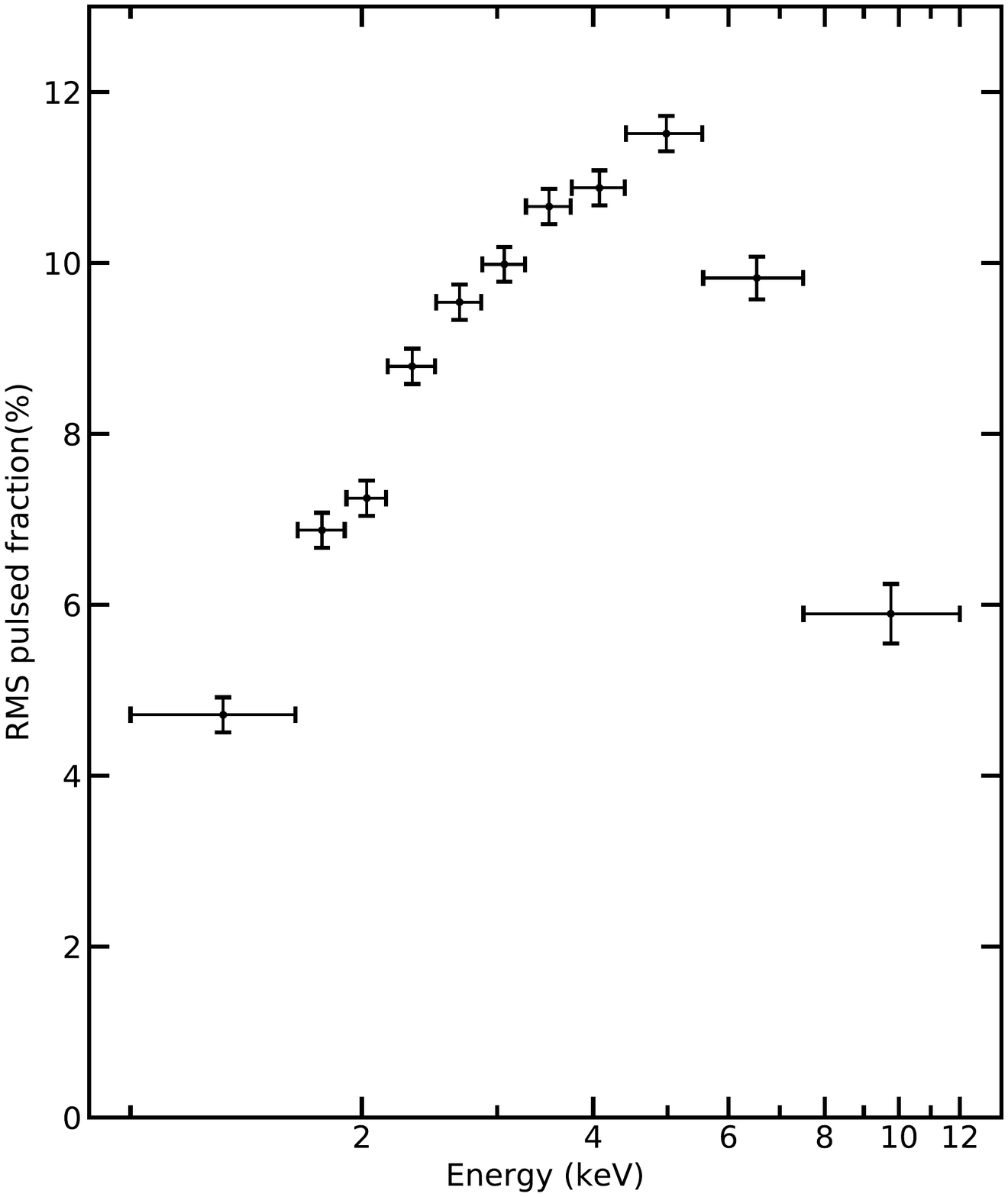}
\caption{Energy dependent rms pulsed fractional amplitude\label{pe}}
\end{figure}

\begin{figure}
  \plotone{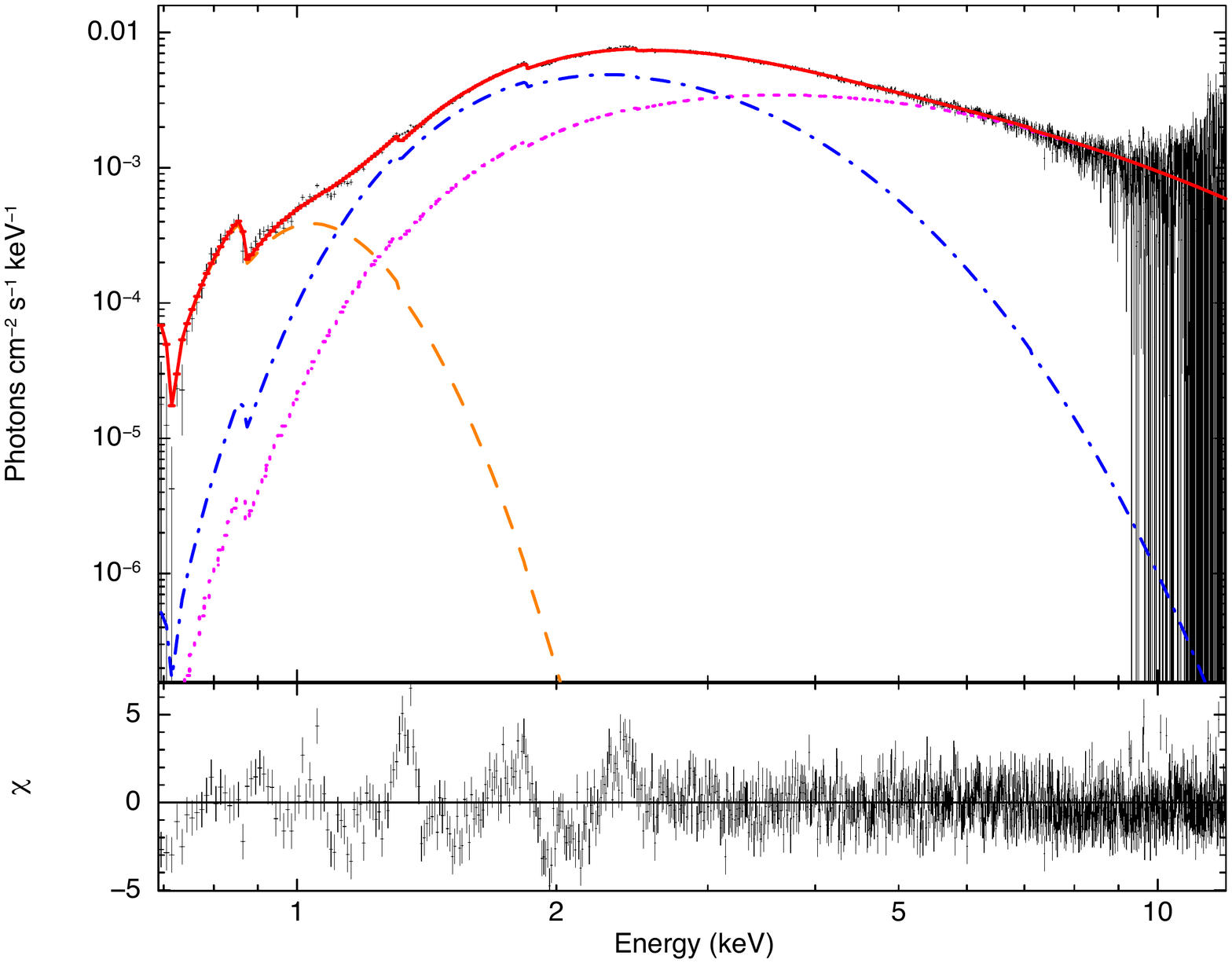}
\caption{Spectrum and the best-fitted model components form {\it NICER} observations. The orange dashed line, blue dash-dotted line, and magenta dotted line represent the disk blackbody (diskbb), blackbody (bbodyrad) and Comptonized (nthcomp) components respectively. The red solid line represents the sum of these three components. The feature below $\sim$2.5 keV in the residuals is likely due to systematic uncertainty of low energy response functions of {\it NICER}.} \label{spec}
\end{figure}

On the other hand, the energy dependent pulse amplitude analysis shows that the rms pulsed fractional amplitudes increases from 1 (4.7\%) to $\sim$5 keV (11.5\%) and then decreases for higher energy bands (Figure~\ref{pe}). The decrease of pulse fractional amplitudes for softer energy bands below $\sim$5 keV are very likely due to the unpulsed component (i.e. the DC term) in the pulse profiles being stronger for the softer energy bands. Adding a multicolor disk blackbody component can significantly improve the spectral fitting, especially for the low energy part. It implies that the the unpulsed component could mainly be due to the thermal emission from the accretion disk. This phenomenon is analogous to the one found in SAX J1808.4-3658 in its 2008 outburst from {\it XMM-Newton} observations~\citep{pat09c}. The pulsed fractional amplitudes largely decrease below 2 keV and a disk blackbody component is also required for acceptable spectral fitting. In fact, a similar phenomenon can be also observed in IGR J00291+5934 during its 2015 outburst for energy bands less than $\sim$2 keV~\citep[see Figure 4 in][]{san17}, although it seems that a disk blackbody component is unnecessary for their spectral fitting. However, we note that the effect of unpulsed emission can be extend up to $\sim$5 keV from the energy dependent pulsed fractional amplitudes but the disk blackbody component from the spectral fitting seems too small to have such a large influence on the unpulsed component of the pulse profile. It is probably due to the imperfect modeling of the Compotonized component, because the electron temperature is fixed at 22 keV. The other possibility is that there is other unpulsed emission from the system, such as thermal emission from the surface of neutron star in the region other than the hot spot. We have attempted to add an additional blackbody component to model the thermal emission from the other part of the surface of neutron star with smaller temperature than the hot spot but failed because these two blackbody components tend to merge together during the fittings, probably due to insufficient spectral resolution to distinguish these two components.

\vspace{5mm}
\facilities{ADS, HEASARC, {\it NICER}}
\software{heasoft (HEASARC 2014), nicerdas (v004), XSPEC~\citep{arn96}}

\clearpage

\clearpage

\clearpage

\clearpage

\clearpage

\clearpage

\clearpage

\clearpage

\clearpage

\clearpage
\begin{table}
\begin{center}
\caption{Spin and Orbital Parameters of IGR J17591-2342 from {\it NICER} Observations\label{para}}
  
\begin{tabular}{lcc}
\\
\tableline\tableline
Parameter & Model 1\tablenotemark{a} & Model 2\tablenotemark{b}\\
\tableline
Orbital period, $P_{orb}$ (s) & 31684.7499(4)  & 31684.7499(4) \\
Projected semimajor axis, $a_x \sin i$ (lt-sec)& 1.227729(3)   &1.227723(3)\\
Ascending node passage, $T_{nod}$ (MJD/TDB)&58345.1719768(3)   &58345.1719769(3)\\
Eccentricity, $e$ &$<1 \times 10^{-5}$  & $<1 \times 10^{-5}$ \\
Spin frequency, $\nu_0$ (Hz)&527.42570061(2)  &527.42570057(2)\\
Spin frequency derivative, $\dot{\nu}$ (Hz s$^{-1}$)& $(-7\pm 1) \times 10^{-14}$  &$(-6 \pm 1) \times 10^{-14}$  \\
Epoch of $\nu_0$, $T_0$ (MJD/TDB) & 58338.0  & 58338.0  \\
$\alpha$ & ------& $0.73 \pm 0.06$ \\
$\chi^2$/d.o.f & 626.94/258 & 475.45/257\\
\tableline
\end{tabular}
\end{center}
\tablenotetext{a}{The flux-dependent timing noise component is not included in the model}
\tablenotetext{b}{The flux-dependent timing noise component is included in the model}
\end{table}

\clearpage
\begin{table}
\begin{center}
\caption{The Best-fitted Spectral Parameters of IGR J17591-2342 from {\it NICER} Observations\label{specpara}}
\begin{tabular}{lc}
\\
\tableline\tableline
Parameter & tbabs(diskbb+bbodyrad+nthcomp) \\
\tableline
$N_H$ ($10^{22}$ cm$^{-2}$) & $2.01^{+0.03}_{-0.02}$  \\
$kT_{in}$ (keV)& $0.081\pm0.002$ \\
Disk flux (1-12 keV) (erg s$^{-1}$ cm$^{-2})$ & $9.20^{+1.04}_{-0.87} \times 10^{-12}$\\
$kT_{BB} (keV) $&$0.64\pm0.02$\\
$Norm_{BB}$\tablenotemark{a} &$57.7^{+3.5}_{-3.0}$\\
BB flux (1-12 keV) (erg s$^{-1}$ cm$^{-2})$ & $9.42^{+0.64}_{-0.83} \times 10^{-11}$\\
BB flux (1-4 keV) (erg s$^{-1}$ cm$^{-2})$ & $8.14^{+0.45}_{-0.61} \times 10^{-11}$\\
$\Gamma$&$2.85^{+0.92}_{-0.50}$\\
$kT_e$(keV)& 22 (fixed)\\
$kT_{seed}$(keV)& $1.48\pm0.19$\\
nthcomp flux (1-12 keV) (erg s$^{-1}$ cm$^{-2})$ & $2.15^{+0.08}_{-0.14} \times 10^{-10}$\\
nthcomp flux (1-4 keV) (erg s$^{-1}$ cm$^{-2}$) & $5.13^{+0.76}_{-0.49} \times 10^{-11}$\\
$\chi^2_{\nu}$ (d.o.f.) & 1.54 (1123)\\ 
\tableline

\end{tabular}
\end{center}
\tablenotetext{a}{The normalization of bbodyrad is presented in a unit of $R^2_{km}/D^2_{10kpc}$ where $R^2_{km}$ is blackbody emission radius in the unit of km and $D^2_{10kpc}$ is the source distance in the unit of 10 kpc.}
\end{table}

\end{document}